\documentclass[12pt,preprint]{aastex}

\shorttitle{Nature of LCBGs in the UDF}
\shortauthors{Noeske et al.}

\begin{document}

\title{Luminous Compact Blue Galaxies up to $z\sim 1$ in the HST Ultra
Deep Field: I. Small galaxies, or blue centers of massive disks?}

\author{K.G. Noeske, D.C. Koo, 
  A.C. Phillips, C.N.A. Willmer, J. Melbourne}
\affil{Lick Observatory, University of California,
    Santa Cruz, CA 95064, USA}
\email{kai@ucolick.org, koo@ucolick.org, cnaw@as.arizona.edu, jmel@ucolick.org, agpaz@astrax.fis.ucm.es,papade@astro.physik.uni-goettingen.de}

\author{A. Gil de Paz}
\affil{Observatories of the Carnegie Institution
of Washington, 813 Santa Barbara Street, Pasadena, CA 91101, USA}

\and

\author{P. Papaderos}
\affil{Institute for Astrophysics, University of  G\"ottingen, Friedrich-Hund-Platz 1, 37077  G\"ottingen, Germany}

\begin{abstract}
We analyze 26 Luminous Compact Blue Galaxies (LCBGs) in the
HST/ACS Ultra Deep Field (UDF) at $z\sim 0.2-1.3$, to determine
whether these are truly small galaxies, or rather bright central
starbursts within existing or forming large disk galaxies.
Surface brightness profiles from UDF images reach fainter than
rest-frame $26.5\,B\,{\rm mag}/\square\arcsec$ even for compact
objects at $z\sim 1$. Most LCBGs show a smaller, brighter component
that is likely star-forming, and an extended, roughly exponential
component with colors suggesting stellar ages $\ga 100$\,Myr to few
Gyr. Scale lengths of the extended components are mostly $\la 2$\,kpc,
$>1.5-2$ times smaller than those of nearby large disk galaxies like
the Milky Way.
Larger, very low surface brightness disks can be excluded down to
faint rest-frame surface brightnesses ($\ga 26\,B\,{\rm
mag}/\square\arcsec$). However, 1 or 2 of the LCBGs are large,
disk-like galaxies that meet LCBG selection criteria due to a bright
central nucleus, possibly a forming bulge.
These results indicate that $\ga90\%$ of high-$z$ LCBGs are small
galaxies that will evolve into small disk galaxies, and low mass
spheroidal or irregular galaxies in the local Universe, assuming
passive evolution and no significant disk growth. The data do not
reveal signs of disk formation around small, H{\small II}-galaxy-like
LCBGs, and do not suggest a simple inside-out growth scenario for
larger LCBGs with a disk-like morphology.
Irregular blue emission in distant LCBGs is relatively extended,
suggesting that nebular emission lines from star-forming regions
sample a major fraction of an LCBG's velocity field.
\end{abstract}

\keywords{galaxies: compact --- galaxies: starburst
--- galaxies: structure --- galaxies: evolution}

\section{Introduction}
\label{intro}

The term ``Luminous Compact Blue Galaxies (LCBGs)'' describes small,
luminous ($\lesssim L_{\star,B}$), high-surface brightness galaxies
with blue optical colors and strong emission lines
\citep{guzman03,garland04,werk04}. Such objects had previously been
classified as e.g. Faint Blue Galaxies, Compact Narrow Emission Line
Galaxies (CNELGs) \citep{koo95,guzman96}, Luminous Compact Galaxies
\citep{hammer01,hammer05}, or Blue Compact Galaxies \citep{pisano01},
with varying selection criteria.

Prior work has indicated that LCBGs are progenitors of different
intermediate - and low mass galaxies in the local Universe that are
brightened by intense star formation (SF) (e.g. \citet{koo95},
\citet{guzman97}). Galaxies undergoing an LCBG phase may contribute
$\sim 45\%$ of the comoving UV-derived SF rate density of the Universe
and $\sim 20\%$ of the field galaxy number density at $z\sim1$
\citep{phillips97,guzman97}, and show the strongest known number
density decline ($\times 10 -100$) from intermediate $z$ $(\sim 0.4 -
1)$ to 0; they are almost absent in the local Universe
\citep{koo94,guzman97,phillips97}. LCBG phases do therefore contribute
sizeably to the evolutionary phenomena observed in the whole galaxy
population to $z=1$ --- the global increase of SF activity
\citep{madau96}, and the luminosity and number density evolution of
blue galaxies \citep{willmer05,faber05}. 

This {\em letter} addresses the controversial question which types of
local galaxies, or which of their subcomponents, experienced the
LCBG phases of massive SF at redshifts $z\ga 0.2$ to $>1$.
\citet{koo95}, \citet{guzman96,guzman98} and \citet{phillips97}
 distinguished smaller LCBGs (half-light radius $r_e \lesssim 3$\,kpc)
 with low velocity dispersion ($\sigma _v \la 65$\,km\,s$^{-1}$) and
 starburst dwarf--like morphologies, and larger, more massive LCBGs
 ($65 \la \sigma _v \la 160$\,km\,s$^{-1}$), more similar to local
 irregular and starburst disk galaxies. They argued that the former
 may fade to ultimately become local low-mass spheroidals/dwarf
 ellipticals while the latter may evolve into local small disks and
 irregulars.
On the other hand, \citet{hammer01,hammer05} and \citet{barton01}
suggested that LCBGs probably represent interaction-induced formation
of bulges in today's massive spiral galaxies, possibly accompanied by
inside-out disk formation \citep{hammer01}. In this scenario, the
apparently small linewidths and sizes of LCBGs do not represent
intrinsic properties of these galaxies: as suggested by \citet{koo95}
and shown by \citet{barton01}, a nuclear starburst in an ordinary
extended disk can skew its effective radius, effective surface
brightness and colors to mimic a blue, compact galaxy. The nuclear
burst's nebular emission would sample only the inner part of the
galaxy's velocity field, and thus lead to an underestimate of its
dynamical mass.  This scenario becomes particularly worrisome with
increasing redshift, where cosmological surface brightness dimming
hampers the detection of low surface brightness (LSB) components, and
nebular emission lines are usually the only indicator of faint
galaxies' kinematics.

To constrain the above scenarios, we present the first structural
study of the extended components in intermediate-$z$ LCBGs, using the
uniquely deep images from the HST/ACS Ultra Deep Field\footnote{56, 56,
144 and 144 orbits of integration time in the $B$(F435W), $V$(F606V),
$i$(F775W) and $z$(F850LP) filters respectively; PI: S. Beckwith,
STScI} (UDF) to search for extended disk components in 26 LCBGs.
Details of this study can be found in an accompanying paper (Noeske et
al. 2005c, in prep.; hereafter Paper II).
Throughout this {\it letter}, we adopt a $\Lambda$CDM cosmology ($H_0
= 70$km\,s$^{-1}$\,Mpc$^{-1}$, $\Omega_M = 0.3, \Omega_{\Lambda} =
0.7$)

\section{Sample selection and data processing}
\label{data}

We adopted similar rest-frame selection criteria to those by
\citet{garland04}, \citet{werk04}, \citet{hoyos04} for local LCBG
samples: (i) blue rest-frame $\bv \leq 0.6$, (ii) average rest-frame
surface brightness within the half-light radius $\mu _e \leq
21\,B$\,mag/$\square\arcsec$), (iii) $M_B \leq -18.5$, and (iv)
half-light radius $r_e \leq 3.5$\,kpc\,.  These somewhat arbitrary
limits include both extremely compact CNELGs, and larger LCBGs, more
comparable to those analyzed by, e.g., \citet{phillips97}.
The galaxies were selected from the UDF SExtractor catalog (Beckwith
et al., in prep.), after computing linear extents and rest-frame
photometry, using the DEEP2 $k-$corrections \citep{willmer05} and
robust spectroscopic \citep{szokoly04,lefevre04,vanzella05,koo05} and
photometric redshifts \citep{wolf04}. After removing doubtful cases
and 2 AGN, this yielded 26 objects at $0.21 < z < 1.25$, $\sim
\twothirds$ at $z> 0.9$. See Fig. \ref{fig3} for examples.

{All LCBG images were analysed through two surface photometry methods:}
(i) 1-d surface brightness profiles (SBPs) were derived using the
morphology-adaptive mask algorithm ``LAZY'' described in
\citet{papaderos02} (``method iv'') and \citet{noeske03}, with
procedures detailed in these papers. {LAZY can treat the irregular
morphologies of LCBGs and is robust at low intensity levels, allowing
to detect and measure, or reject, large LSB structures.} The resulting
SBPs typically showed approximately exponential, moderately extended
components (see Section \ref{results}) which we fitted by exponential
laws outside the brighter central excesses (see \citet{noeske03}).
{(ii) PSF effects on measured structural parameters of the extended
components are non-negligible. For brighter components with roughly
elliptical isophotes, PSF treatment is provided by the GALFIT code
\citep{peng02}.} We decomposed the LCBGs into two exponential
components.  Obviously unphysical fits due to very irregular
morphologies were rejected, as well as fitted extended components
fainter than an empirical reliability limit of 25 AB mag in $i$ and
$z$ .

Comparisons between between LAZY and GALFIT, and reliability
assessments are detailed in Paper II. Exponential scale lengths
($R_s$) from LAZY are typically overestimated for small objects ($R_s
\la 1.5$\,kpc) by a factor of $\la$1.3, few up to $\sim 2$, while for
larger scale lengths, GALFIT can underestimate $R_s$ by a factor $\la$
1.3 (see Fig. \ref{fig4}). {Both methods hence bracket the true scale
lengths.} Examples of SBPs are shown in Fig. \ref{fig3}. In most cases,
the rest-frame $B$ band SBPs reached beyond the rest-frame Holmberg
radius ($26.5\,B\,{\rm mag}/\square\arcsec$) even for compact objects
at $z\sim 1$ (see object UDF0901 in Fig. \ref{fig3}).

\section{Results}
\label{results}

\subsection{LCBGs: star formation within more extended, evolved galaxies}

The SBPs of most LCBGs display a moderately extended, roughly
exponential component, corresponding to a mostly fairly regular outer
component in the images (Fig. \ref{fig3}). At smaller radii, the SBPs
show brighter, smaller components, in excess of the extended
exponentials. These range from bright nuclei to extended emission over
a large part of the galaxy, and reflect the irregular blue emission
seen in the images, i.e. probably the ongoing SF.
For larger LCBGs, this structure has previously been reported
\citep{koo95,phillips97,guzman98,hammer01}. For smaller LCBGs, our
current UDF dataset {verifies} what the data by \citet{guzman98}
suggested: also these objects, similar to local H{\small II}
galaxies or distant CNELGs, have roughly exponential stellar
components that pre-date the ongoing SF.

{Rest-frame colors of the extended components are $-0.3\la
\ub\la 0.3$ and $0.3\la \bv\la 0.9$ for $\sim 90\%$ of the LCBGs,
on average $\sim 0.2$ mag redder than the SF excesses.  For these
colors, simple stellar population models for up to solar metallicity
\citep{anders03} yield minimum stellar ages of $\ga 100$\,Myr to
several Gyr. More extended SF histories (e.g. \citet{bicker04}) lead
to higher ages. If extinction is significant in the extended
components, outside strong SF, these age limits will decrease.}

We will refer to 'extended components' rather than 'disks' as we lack
resolved kinematic data for most objects. Morphologies are often
disk-like for larger LCBGs (see \ref{fig3}), but ambiguous for small
objects which could be either spheroids or disks.

\subsection{Structure of the extended components}

Figure \ref{fig4} compares the extended components of the LCBGs to
samples of {nearby disk galaxies and dwarf galaxies} with
exponential SBPs. The disk sample by \citet{lu98} was chosen because
of its completeness of local field disk scalelengths, and the
UMa-cluster sample by \citet{tully96} was added to include disks from
higher density environments.

It is evident that the extended components of almost all LCBGs have
small scale lengths ($R_s\la2$\,kpc): at any given luminosity, they
are comparable to the local disk galaxies with the smallest
$R_s$. This result is robust against uncertainties of $R_s$. The less
luminous exponential components are also compatible with those of
relatively compact, luminous local dwarf galaxies (compact dEs, or
stellar hosts of BCDs).
The only two exceptions are UDF0900 (the LCBG with the largest $R_s$
in Fig. \ref{fig4}, see also Fig. \ref{fig3}), which shows a large LSB
component ($\mu _{0,B}\approx 22.9{\rm mag}/\square\arcsec$,
$R_s=4.4$\,kpc) with an exponential SBP but no spiral features, {and a
galaxy that may have a truncated disk with a small outer, but larger
inner scalelength} (Fig. \ref{fig4})\footnote{The second-largest
LAZY-derived scalelength in Fig. \ref{fig4} is affected by PSF wings
from a bright nucleus and in fact smaller; see the connected GALFIT
data point.}  Both galaxies have bright central regions that led to
their classification as LCBGs.

\section{Discussion and Conclusions} 

\begin{figure*}
\epsscale{1.2}
\centerline{
\includegraphics[height=1.8in]{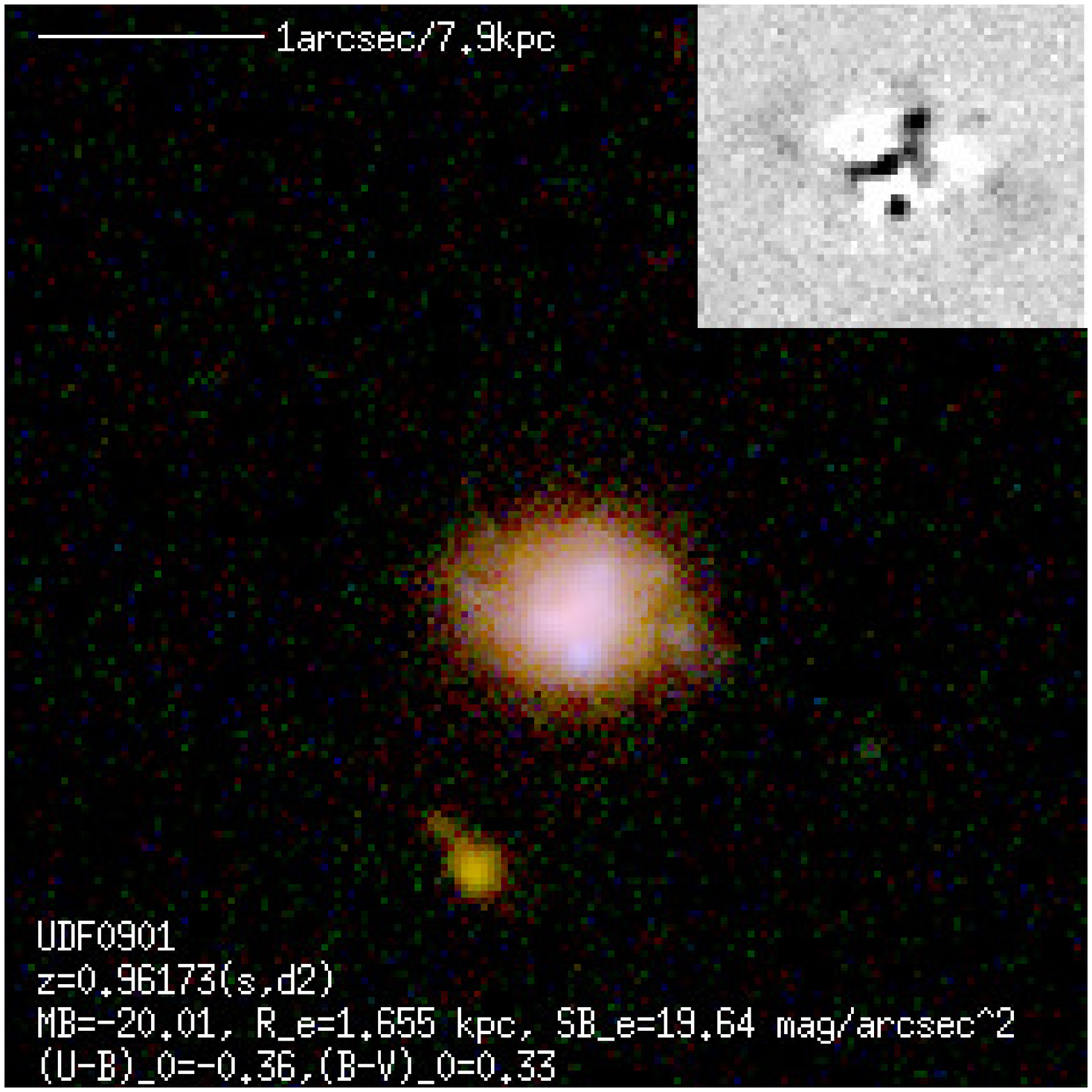}
\includegraphics[height=1.8in]{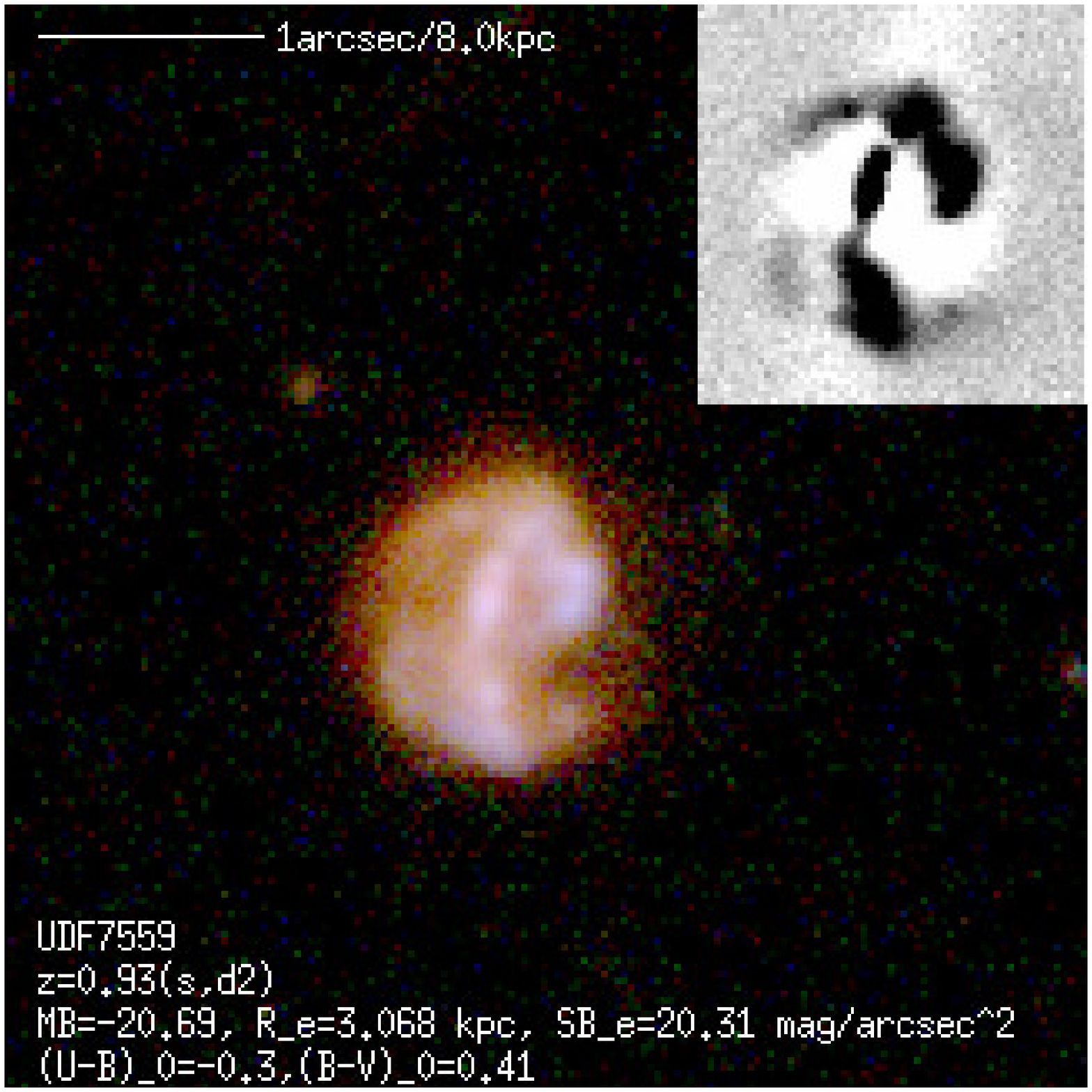}
\includegraphics[height=1.8in]{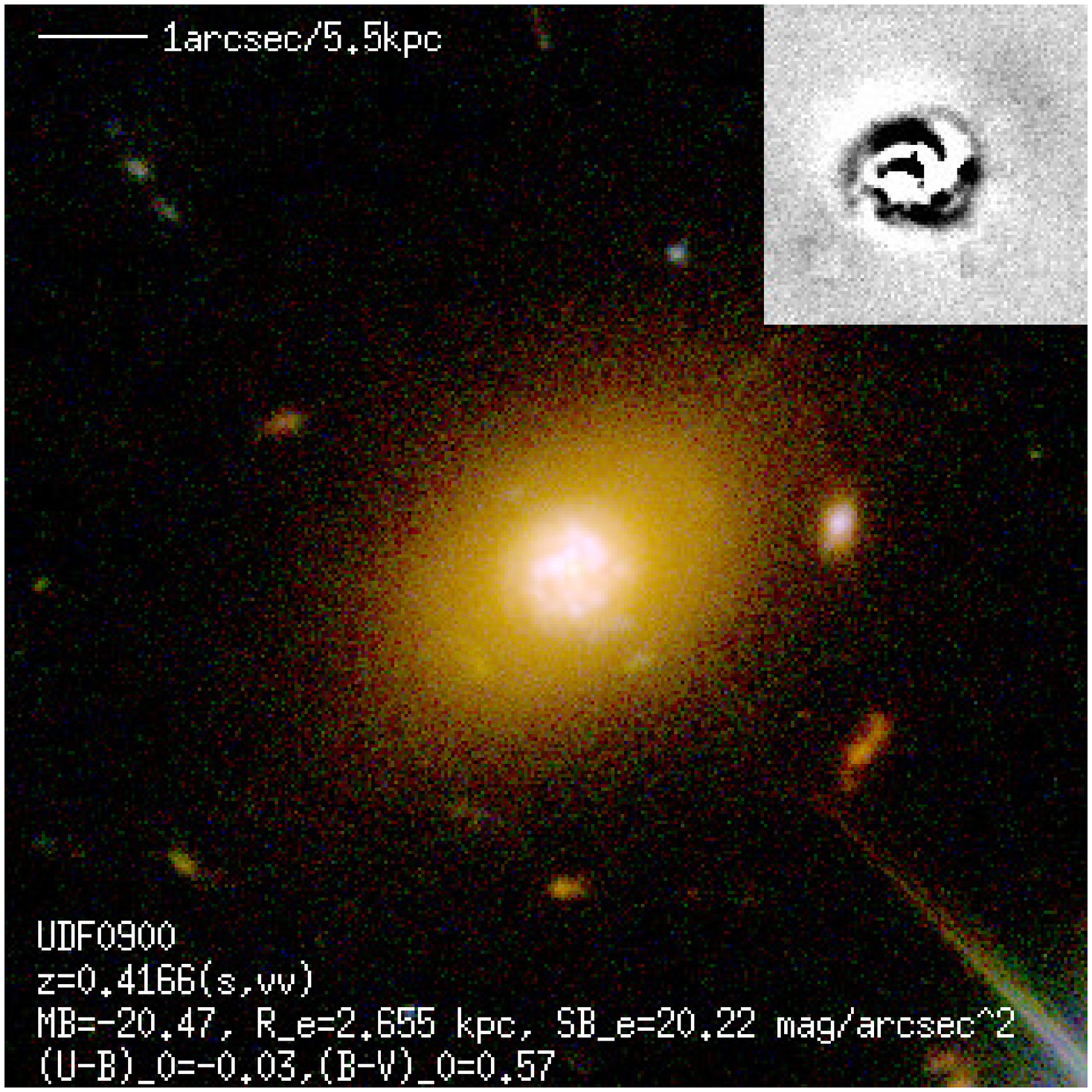}
}
\centerline{
\includegraphics[width=2.33in]{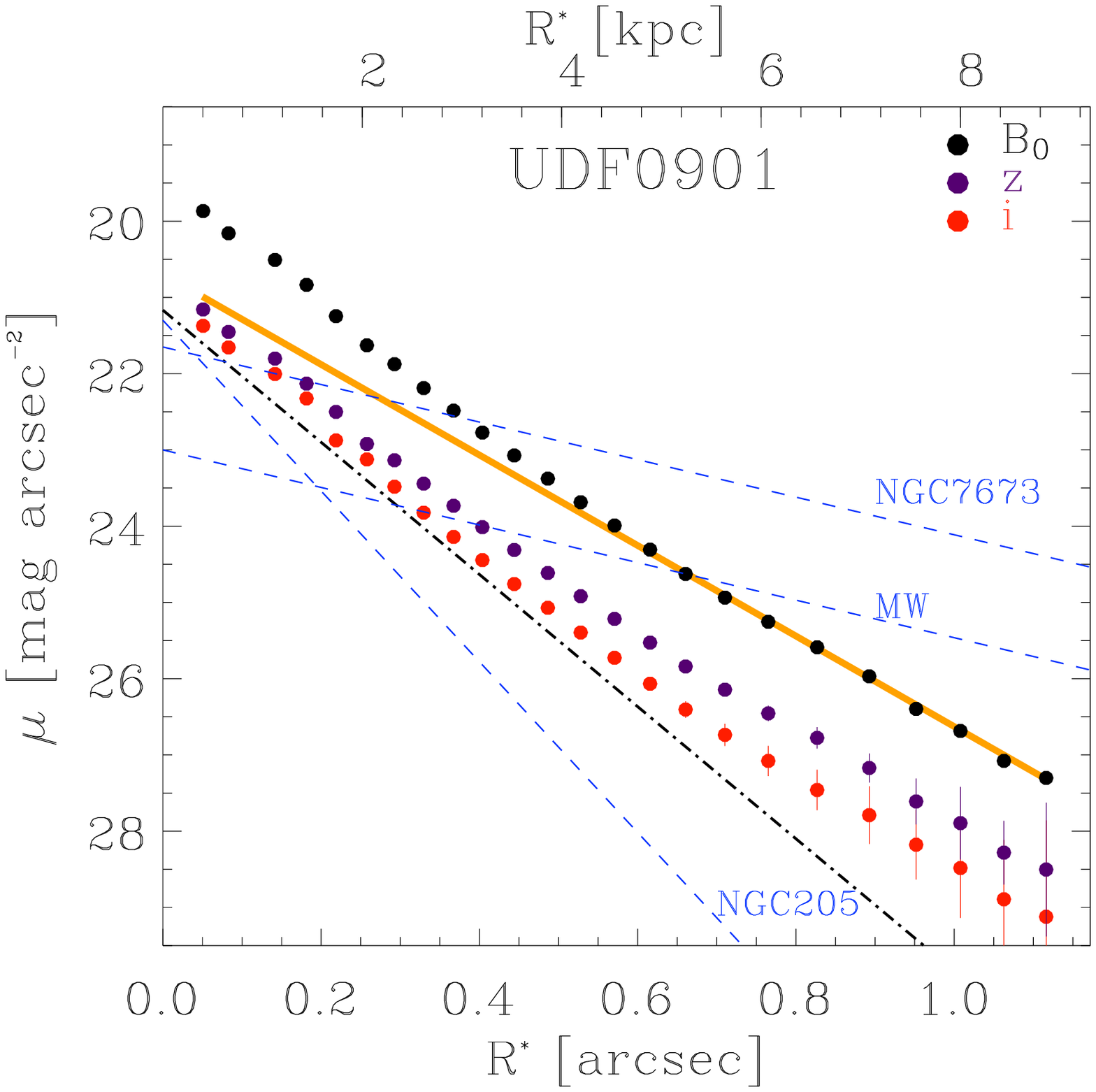}\hspace*{-4mm}
\includegraphics[width=2.33in]{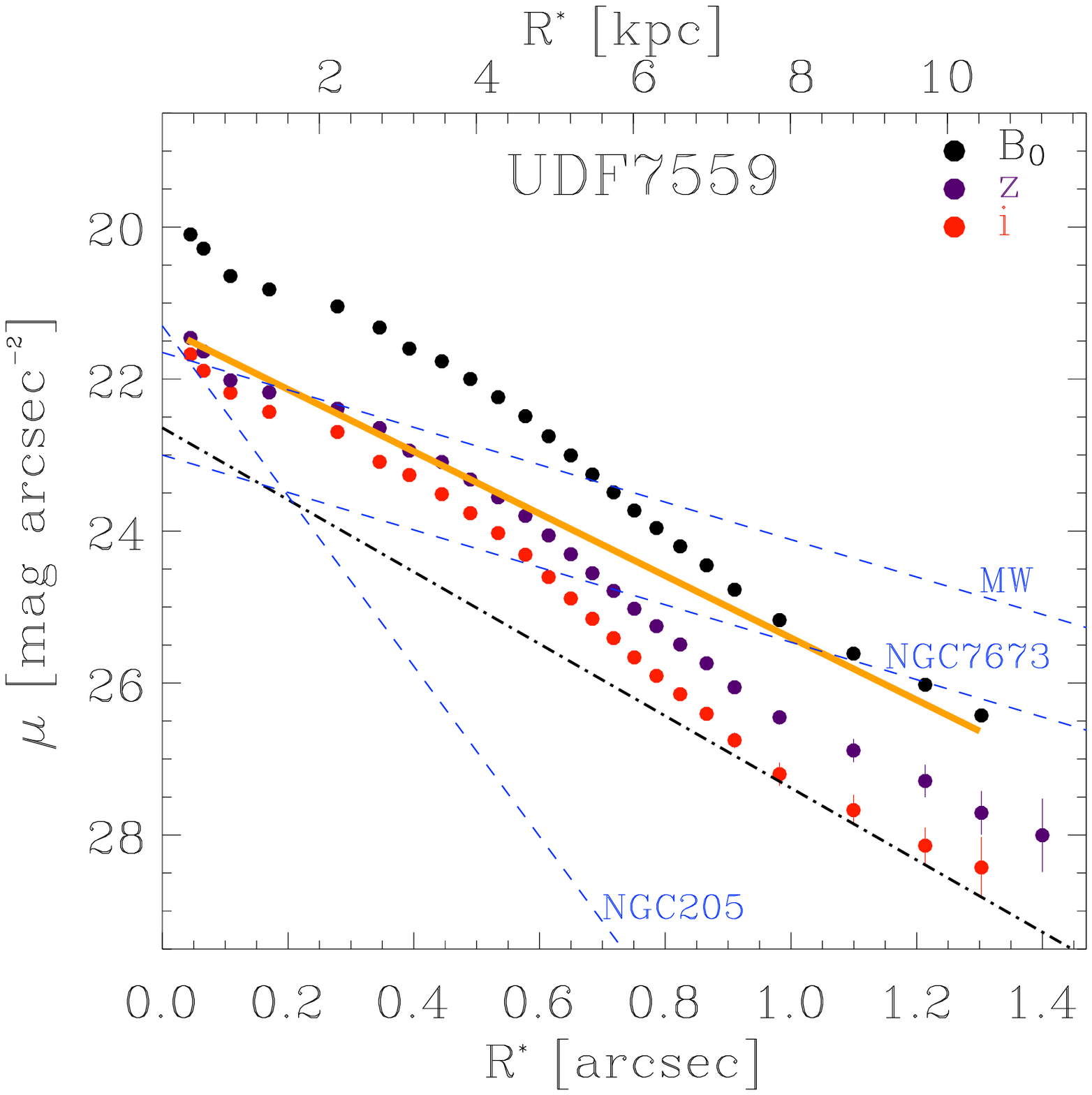}\hspace*{-4mm}
\includegraphics[width=2.33in]{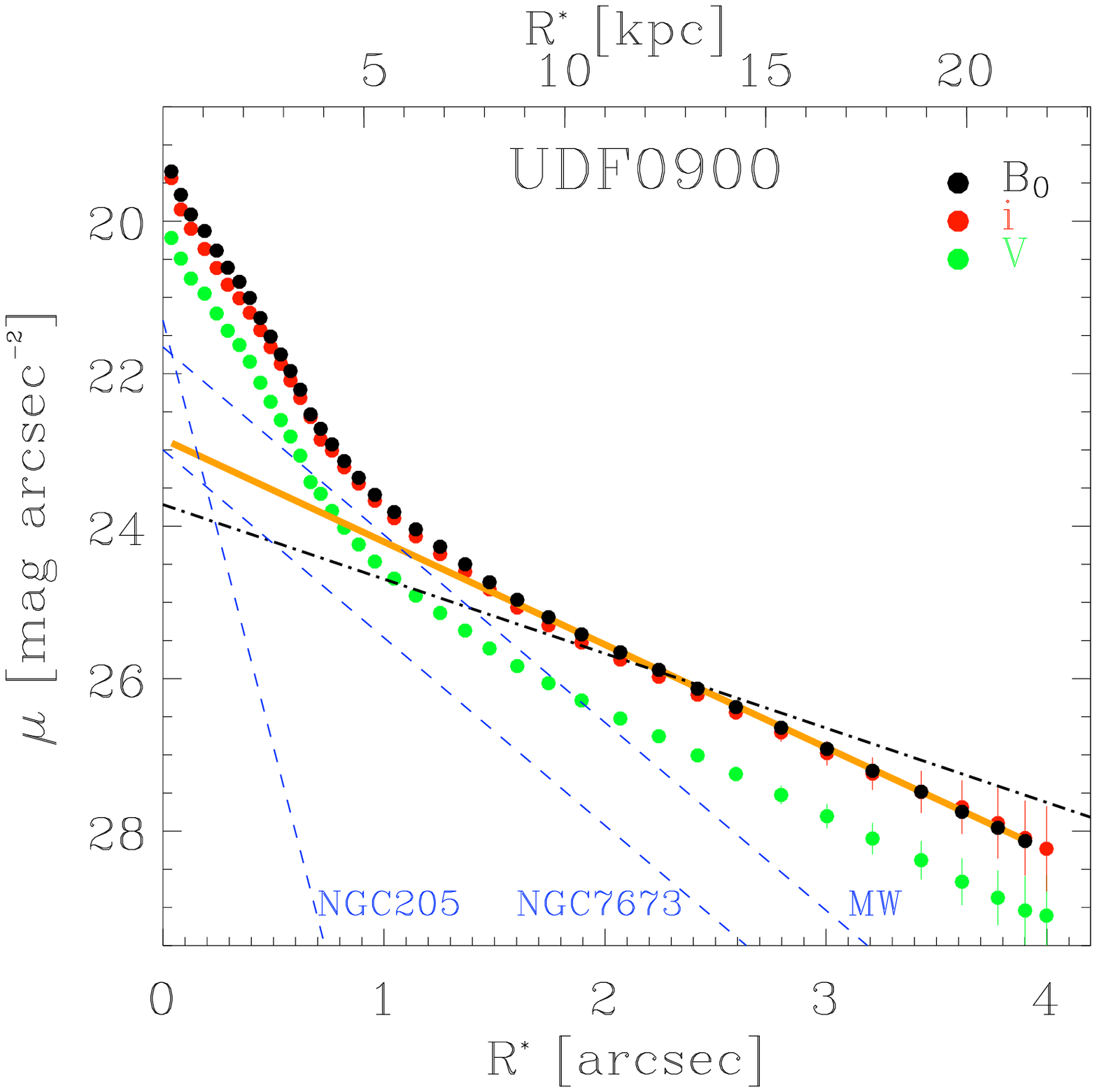}
}
\caption{\label{fig3} {\bf Three-color images:} from HST $B,V,i$
images, using non-linear, non-saturating intensity scaling
\citep{lupton04}. UDF0901 ($\sigma _v$ = 51.6 km/s) and UDF7559
($\sigma _v$ = 109.5 km/s) are examples for typical LCBGs- compact,
narrow emission line (CNELGs) and more extended, broader-line
objects. UDF0900 is one of two extended, low-surface brightness (LSB)
galaxies with a bright, compact nucleus. {\bf Gray insets:}
small-scale $i$ band residuals after subtracting smooth GALFIT
models. Spatial scaling is equal to the three-color images.  {\bf
Surface brightness profiles:} Colored dots ($V,i,z$) show the observed
profiles closest in wavelength to the rest-frame $B$ band. Black dots
($B_0$) denote the rest-frame $B_{\rm Vega}$ profile, $k$-corrected
and corrected for cosmological surface brightness dimming.  {\bf
Dashed blue lines:} The Milky Way (MW) disk, a LSB disk in a nearby
LCBG (NGC 7673), and the dE NGC 205 \citep{choi02} in rest-frame $B$,
for comparison. Note the surface brightness limits of the rest-frame
$B$ SBPs $>26 {\rm mag}/\square\arcsec$, and the detectability of
large disks at redshifts $z\sim1$ . {\bf Thick orange lines} show fits
to the extended exponential components in rest-frame $B$ LAZY
profiles. {\bf Dot-dashed lines} give the extended exponential
component yielded by GALFIT decompositions in the $i$ band, to
illustrate PSF effects on LAZY-derived profiles. }
\end{figure*}

\begin{figure}
\epsscale{.80}
\plotone{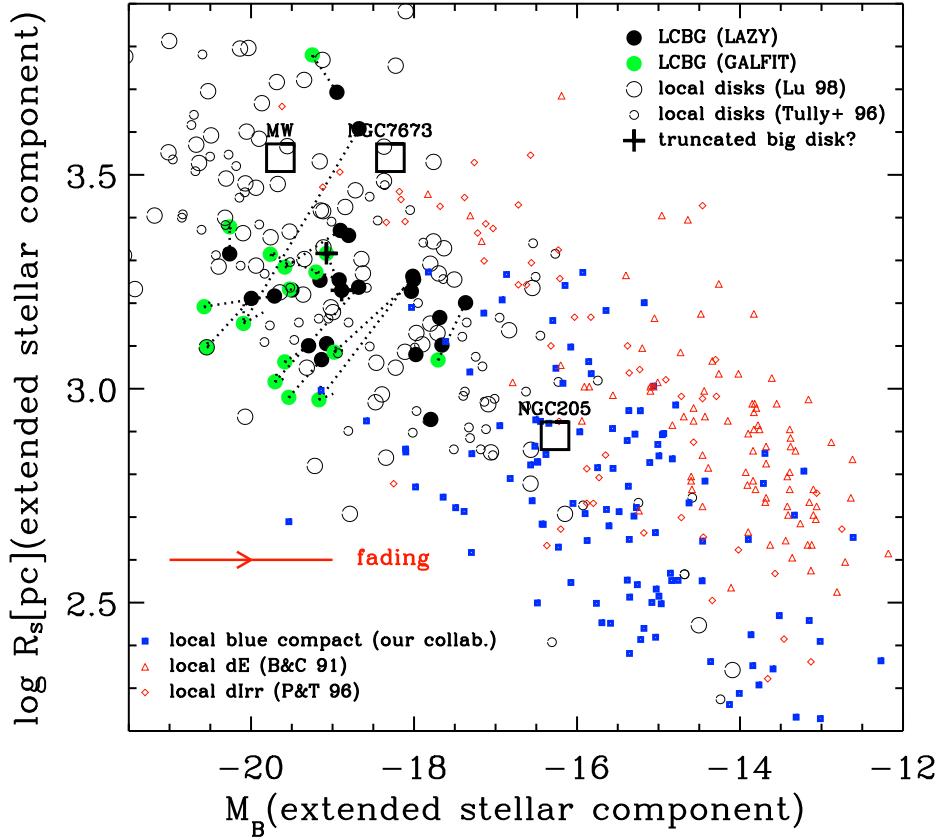}
\caption{ \label{fig4} Exponential scale length $R_s$ vs. absolute $B$
band magnitude for the extended exponential components in LCBGs. Black
filled circles denote LAZY, green ones GALFIT decompositions (see
Section \ref{data}). LAZY scalelengths are inclination-corrected,
assuming that the extended components are inclined disks, i.e. are
upper limits for spheroids. Open circles: local disk galaxy samples
from the UMa cluster \citep{tully96} and from field environments in
the local supercluster \citep{lu98}. Blue squares: stellar host
galaxies of Blue Compact Dwarfs (\citet{gildepaz04} and references
therein), red lozenges: dwarf irregulars \citep{patterson96}, red
triangles: dwarf ellipticals \citep{binggeli91}. Open boxes: the dwarf
elliptical NGC 205 \citet{choi02}, and the disks of the MW (3.5\,kpc,
\citet{devauc78}) and the the nearby LCBG candidate NGC7673 (see
\citet{pisano01}; SBP derived in this work, see Paper II). The arrow
shows the effect of fading.}
\end{figure}

This structural study of their extended components reveals that $\sim
90\%$ of LCBGs at $z\sim 0.2-1.3$ are truly small galaxies. For these,
suspected large disks with $R_s$ similar to the MW, or the extended
low-surface brightness component found in a local LCBG (NGC\,7673),
can be ruled out down to surface brightnesses $\ga 26 \,B\,{\rm
mag}/\square\arcsec$ (cf. Fig. \ref{fig3}).  {The discovery of 1,
possibly 2 LCBGs ($\la 10\%$) that are large, disk-like galaxies with
bright nuclei supports the scenario of some LCBGs at higher $z$ being
nuclear starbursts, possibly bulge formation, in large disks.}

The available scale lengths help to constrain {the passive,
post-starburst evolution scenario for LCBGs. Fading of the smaller,
brighter SF component will affect the {\it overall} photometric
structure of an LCBG. The older extended components alone will however
likely fade more homogeneously and thus largely maintain their scale
lengths.} The extended components could fade by several mag, the
bluest in principle up to 5 $B$ mag from $z=1$ to $0$ if they were
$\sim$100 Myr old simple stellar populations (see above). However, the
scale lengths suggest that LCBGs evolve into local small disks, and
different types of {larger dwarf and low mass galaxies}, assuming
that their subsequent evolution does not involve strong disk
growth\footnote{For low-mass LCBGs similar to Blue Compact Dwarf
galaxies, also energy input from strong SF could affect their stellar
mass distribution \citep{papaderos96}.}.

Subsequent evolution into local large disks of at least the MW scale
length would require an LCBG's extended stellar component to grow
substantially, by a factor $\ga 1.5$, typically $>2$. The data cannot
exclude, but neither evidence, such ongoing growth:
For larger LCBGs with a disk-like morphology, we find no evidence of a
simple inside-out growth such as bluening at large radii. For these
galaxies, local disks with small scalelengths provide a possible
descendant population, so that the substantial disk growth that
\citet{hammer01,hammer05} propose for similar and somewhat larger
LCBGs may not be required\footnote{Note that the selection criteria by
Hammer et al. and this work are not fully comparable; their LCBG
criteria may favor progenitors of larger local galaxies than our
sample.}. From their linewidths, sizes, and stellar masses
(\cite{guzman03}), disk-like LCBGs could well be LCBG phases of local
intermediate-mass ($M_{\star}\sim 10^{10}M_{\sun}$) disk galaxies, in
agreement with the finding that such galaxies experienced substantial
SF since $z\sim 1$ \citep{heavens04,bell05,hammer05}.
Around small, H{\small II}-galaxy like LCBGs, we do not find signs of
forming or pre-existing big disks. Their morphologies and sizes
make these galaxies candidate progenitors of small, low mass galaxies
in the local Universe, such as low mass spheroidals or irregulars.
{Resolved kinematic data will be important to constrain such scenarios.

We finally note that the blue, irregular emission in most LCBGs that
are not large disks extends out to $\ga 1.5$ to $2$ scale lengths of
the extended component (grey insets in Fig. \ref{fig3}). This
fractional extent is similar to nearby LCBGs \citep{papaderos96},
where this blue emission is the locus of the ongoing SF
and nebular emission. It appears plausible that irregular emission
traces nebular emission also in distant LCBGs. If so, then nebular
emission will sample a sufficient fraction of their velocity field to
provide valid mass estimates: \citet{pisano01} showed that nebular
emission line kinematics in nearby LCBGs trace neutral gas kinematics
with moderate correction factors ($\sim 0.7$). This lends support to
distant LCBGs being mostly low- to intermediate mass galaxies.}


{In summary, most LCBGs at intermediate $z$ show brighter,
irregular, likely star-forming emission within more extended, regular
components with approximately exponential intensity distributions and
minimum stellar ages $\ga$ 100 Myr. Most extended components have
scale lengths by factors $>1.5 - 2$ smaller than local large disks
such as the Milky Way, while 1 or 2 of 26 LCBGs are larger, disk-like
galaxies with bright nuclei. This suggests that $\sim 90\%$ of LCBGs
are progenitors of small disks, irregulars or low-mass spheroidals in
the local Universe; $\sim 10\%$ may represent bulge formation within
massive disks.}

\acknowledgments We wish to thank Dr. M. Bershady for the WIYN
$R$-band image of NGC7673. Research by DCK and KGN was partly funded
by the NSF grant AST-0071198 and the HST grants GO-09126.02-A and
AR-10321.02-A\,. We thank the referees for their valuable comments on
this paper, and Dr. B. Holden for helpful advice on ACS PSF
peculiarities.


\clearpage

\clearpage

\end{document}